\documentclass[12pt]{iopart}
\usepackage{amssymb,graphicx,cite}
\begin{document}
\title[Conjunction of $\gamma$-rigid and $\gamma$-stable collective motion...]{Conjunction of $\gamma$-rigid and $\gamma$-stable collective motion in the critical point of the phase transition from spherical to deformed nuclear shapes.}
\author{R. Budaca$^{1}$ and A. I. Budaca$^{1}$}
\address{$^{1}$Horia Hulubei National Institute of Physics and Nuclear Engineering, RO-077125 Bucharest-Magurele, Romania}
\eads{\mailto{rbudaca@theory.nipne.ro}, \mailto{abudaca@theory.nipne.ro}}
\begin{abstract}
Based on the competition between $\gamma$-stable and $\gamma$-rigid collective motions mediated by a rigidity parameter, a two-parameter exactly separable version of the Bohr Hamiltonian is proposed. The $\gamma$-stable part of the Hamiltonian is restricted to stiff oscillations around the $\gamma$ value of the rigid motion. The separated potential for $\beta$ and $\gamma$ shape variables is chosen such that in the lower limit of this parameter, the model recovers exactly the ES-$X(5)$ model, while in the upper limit it tends to the prolate $\gamma$-rigid solution $X(3)$. The combined effect of the rigidity and stiffness parameters on the energy spectrum and wave function is duly investigated. Numerical results are given for few nuclei showing such ambiguous behaviour.
\end{abstract}


\vspace{2pc}

\submitto{\jpg}

\section{Introduction}

Since its formulation \cite{Bohr1,Bohr2}, the Bohr-Mottelson model remained the indispensable tool for interpretation of the collective phenomena in even-even nuclei, with numerous solutions being constantly proposed by considering different shapes and approximations for the potential energy $V(\beta,\gamma)$ \cite{Fortunato1,Casten,Cejnar}. These simple solutions generated in turn other more involved approaches by considering octupole degrees of freedom \cite{Rohozinski,Bizzeti,Bon1} and deformation-dependent mass terms \cite{Bon2,Bon3}, or more simple ones by imposing $\gamma$-rigidity \cite{Bon4,Bon5}. The formalism of the Bohr-Mottelson model recently inspired similar approaches for the definition of collective Hamiltonians related to such phenomena like the chiral modes \cite{Chen1} or the wobbling excitations \cite{Chen2}. The reformulation of the Bohr Hamiltonian in six dimensions \cite{Georg} is another recent noteworthy addition to the geometrical description of collective motion.

The parameter independent solutions of the Bohr Hamiltonian are especially useful in the sense that they serve as reference points for the energy spectra and electromagnetic properties of nuclei. It was found that this is also the case for the critical point symmetries $E(5)$ \cite{Iachello0} and $X(5)$ \cite{Iachello1} describing the shape phase transitions from vibrational to $\gamma$-unstable and respectively to the axially deformed nuclei. The $E(5)$ solution is exact, while $X(5)$ employs two approximations, one related to the separation of variables and the other based on the small angles for the $\gamma$ shape variable. The domains of validity for these approximations were found to be incongruous when the Hamiltonian is numerically diagonalized \cite{Caprio} in the framework of the recently developed Algebraic Collective Model \cite{Rowe1,Rowe2,Rowe3}. Also, due to its structure, $X(5)$ is parameter free only for the ground and $\beta$ bands, while the $\gamma$ band is determined by the stiffness of the $\gamma$ oscillations. All these shortcomings were commented on all accounts in \cite{BonX5ES}, where an exactly separable version of the $X(5)$ model denoted ES-$X(5)$, was proposed. As in this case the $\beta$ and $\gamma$ degrees of freedom are treated on the same footing, the model will no longer be parameter independent, the whole energy spectrum depending on the $\gamma$ stiffness parameter. The $\gamma$-rigid version of this model, called $X(3)$ \cite{Bon5}, however will be parameter independent. Although the $\gamma$-rigidity hypothesis is somehow a crude one, it provides simple approaches with successful reproduction of the relevant experimental data \cite{Bon4,Bon5,Davydov,Budaca1,Budaca2,Buganu,Zhang}. Moreover, an unexpected similarity between the $\beta$ excited bands of $X(5)$ and $X(3)$ solutions was found, which inevitably address the question about the importance of rigidity in explaining the critical collective phenomena.

In the present work we propose a simple exactly separable model for the competition between the $\gamma$-rigid and $\gamma$-stable collective motion in the phase transition between spherical and deformed shapes. The coupling of the two types of $\beta$ vibration is achieved by introducing a control parameter measuring the degree of the system's $\gamma$-rigidity in an Ising type Hamiltonian \cite{Casten}. The separation of variables is achieved by considering a potential of the form $u(\beta)+u(\gamma)/\beta^{2}$ \cite{Fortunato1,BonX5ES,Wilets,Fortunato2,Fortunato3} adapted to the current problem. Matching the two competing excitations, the $\gamma$ potential is chosen to be a harmonic oscillator centered in $\gamma=0$, which is consistent with the prolate $\gamma$-rigid part of the problem. As the purpose of this paper is to investigate the $\gamma$ rigidity near a critical point which is commonly identified with a very flat potential, $u(\beta)$ is taken to be an infinite square well. Although the separation of the potential leads to independent analytical descriptions of the $\beta$ and $\gamma$ degrees of freedom, it contravenes the Liquid Drop Model prescription for the potential energy \cite{Bohr3}. However, the chosen potential shape can be simulated by considering higher order anharmonic terms in the generalized collective geometrical model \cite{Gneus}.

The energy spectrum and $E2$ transition probabilities resulting from the proposed model depend up to an overall scaling factor on two parameters, namely the rigidity and the stiffness of the $\gamma$ vibrations. Their separate influence on the model's characteristics is investigated through numerical applications. The experimental realization of the model is found in few rare earth nuclei around $N=96$.

\renewcommand{\theequation}{2.\arabic{equation}}
\section{Interplay between $\gamma$-stable and $\gamma$-rigid collective motion}
\label{sec:2}
The kinetic energy operator associated to a prolate $\gamma$-rigid nucleus is \cite{Bon5,Budaca1,Budaca2}:
\begin{equation}
T_{r}=-\frac{\hbar^{2}}{2B}\left[\frac{1}{\beta^{2}}\frac{\partial}{\partial{\beta}}\beta^{2}\frac{\partial}{\partial{\beta}}-\frac{\mathbf{Q}^{2}}{3\beta^{2}}\right],
\label{Tr}
\end{equation}
while in the $\gamma$-stable case it is expressed as
\begin{eqnarray}
T_{s}&=&-\frac{\hbar^{2}}{2B}\left[\frac{1}{\beta^{4}}\frac{\partial}{\partial{\beta}}\beta^{4}\frac{\partial}{\partial{\beta}}+\frac{1}{\beta^{2}\sin{3\gamma}}\frac{\partial}{\partial\gamma}\sin{3\gamma}\frac{\partial}{\partial\gamma}\right.\nonumber\\
&&\left.-\frac{1}{4\beta^{2}}\sum_{k=1}^{3}\frac{Q_{k}^{2}}{\sin^{2}{\left(\gamma-\frac{2}{3}\pi k\right)}}\right],
\label{Ts}
\end{eqnarray}
where $\mathbf{Q}$ is the angular momentum operator from the intrinsic frame of reference with the corresponding components $Q_{k}(k=1,2,3)$, while $B$ is the mass parameter. The interplay between $\gamma$-stable and $\gamma$-rigid collective motion is achieved by considering the Hamiltonian:
\begin{equation}
H=\chi T_{r}+(1-\chi)T_{s}+V(\beta,\gamma),
\label{Ht}
\end{equation}
where $0\leqslant\chi<1$ is a parameter which indicates the degree of the system's rigidity against $\gamma$ vibrations. Note that the prolate $\gamma$-rigid limit $\chi=1$ is avoided in order to preserve the geometry of the curvilinear space.

In order to achieve an exact separation of the $\beta$ variable from the $\gamma$-angular ones, the reduced potential is considered of the form similar to that used in \cite{Fortunato1,BonX5ES,Wilets,Fortunato2,Fortunato3} but adapted for the present problem:
\begin{equation}
v(\beta,\gamma)=\frac{2B}{\hbar^{2}}V(\beta,\gamma)=u(\beta)+(1-\chi)\frac{u(\gamma)}{\beta^{2}}.
\label{pot}
\end{equation}
Factorizing the total wave function as $\Psi(\beta,\gamma,\Omega)=\xi(\beta)\varphi(\gamma,\Omega)$, the associated Schr\"{o}dinger equation is separated in two parts:
\begin{equation}
\left[-\frac{\partial^{2}}{\partial{\beta^{2}}}-\frac{2(2-\chi)}{\beta}\frac{\partial}{\partial{\beta}}+\frac{W}{\beta^{2}}+u(\beta)\right]\xi(\beta)=\epsilon\xi(\beta),
\label{b}
\end{equation}
and
\begin{eqnarray}
\left[(1-\chi)\left(-\frac{1}{\sin{3\gamma}}\frac{\partial}{\partial\gamma}\sin{3\gamma}\frac{\partial}{\partial\gamma}+\sum_{k=1}^{3}\frac{Q_{k}^{2}}{4\sin^{2}{\left(\gamma-\frac{2}{3}\pi k\right)}}\right)\right.&&\nonumber\\
+\frac{\chi}{3}\mathbf{Q}^{2}+(1-\chi)u(\gamma)\Bigg]\varphi(\gamma,\Omega)=W\varphi(\gamma,\Omega),&&
\label{ga}
\end{eqnarray}
where $\epsilon=\frac{2B}{\hbar^{2}}E$. As the present model is based on the hypothesis of small oscillations around $\gamma=0$ which cease at $\chi=1$, the rotational term of the last equation is approximated as in \cite{Iachello1}:
\begin{equation}
\sum_{k=1}^{3}\frac{Q_{k}^{2}}{\sin^{2}{\left(\gamma-\frac{2}{3}\pi k\right)}}\approx\frac{4}{3}\mathbf{Q}^{2}+Q_{3}^{2}\left(\frac{1}{\sin^{2}{\gamma}}-\frac{4}{3}\right).
\end{equation}
Contrary to the $\gamma$-unstable case \cite{Bes}, the $\gamma$ and angular variables can also be separated by considering the product state $\varphi(\gamma,\Omega)=\eta(\gamma)D^{L}_{MK}(\Omega)$. Here $D^{L}_{MK}(\Omega)$
are the Wigner functions associated with the total angular momentum $L$ and its projections on the body-fixed and laboratory-fixed $z$ axis, $M$ and $K$, respectively. Averaging the approximated equation (\ref{ga}) on this product state one obtains the following equation for the $\gamma$ shape variable:
\begin{equation}
\left[-\frac{1}{\sin{3\gamma}}\frac{\partial}{\partial\gamma}\sin{3\gamma}\frac{\partial}{\partial\gamma}+\frac{K^{2}}{4\sin^{2}{\gamma}}+u(\gamma)\right]\eta(\gamma)=\epsilon_{\gamma}\eta(\gamma),
\label{ecg}
\end{equation}
with
\begin{equation}
\epsilon_{\gamma}=\frac{1}{1-\chi}\left(W-\frac{L(L+1)-(1-\chi)K^{2}}{3}\right).
\end{equation}
For small oscillations around $\gamma=0$, the "frozen" value of the $\gamma$-rigid part of the Hamiltonian, it is natural to adopt a harmonic oscillator form for the $\gamma$ potential,
\begin{equation}
u(\gamma)=(3a)^{2}\frac{\gamma^{2}}{2},
\label{gamapot}
\end{equation}
where the parameter $a$ defining the string constant of the oscillator is usually referred to as the stiffness of the $\gamma$ vibrations. Then (\ref{ecg}) can be treated as in \cite{Iachello1,BonX5ES}, {\it i.e.} by applying a harmonic approximation for the trigonometric functions around $\gamma=0$. This leads to a differential equation for $\gamma$ which resembles the radial equation for a two dimensional harmonic oscillator:
\begin{equation}
\left[-\frac{1}{\gamma}\frac{\partial}{\partial\gamma}\gamma\frac{\partial}{\partial\gamma}+\left(\frac{K}{2}\right)^{2}\frac{1}{\gamma^{2}}+(3a)^{2}\frac{\gamma^{2}}{2}\right]\eta(\gamma)=\epsilon_{\gamma}\eta(\gamma).
\label{ecga}
\end{equation}
The solutions are readily obtained in terms of the Laguerre polynomials \cite{Iachello1}:
\begin{equation}
\eta_{n_{\gamma},|K|}(\gamma)=N_{n,|K|}\gamma^{|K/2|}\exp{\left(-3a\frac{\gamma^{2}}{2}\right)}L_{n}^{|K/2|}(3a\gamma^{2}),
\label{fg}
\end{equation}
where $N_{n,|K|}$ is a normalization constant and $n=(n_{\gamma}-|K/2|)/2$. The corresponding eigenvalues are
\begin{equation}
\epsilon_{\gamma}=3a(n_{\gamma}+1),\,\,n_{\gamma}=0,1,2,...,
\end{equation}
with $K=0,\pm 2n_{\gamma}$ for $n_{\gamma}$ even and $K=\pm 2n_{\gamma}$ for $n_{\gamma}$ odd, respectively. At this point it would be opportune to comment the implications of choosing such a $\gamma$ potential. First of all it loses the symmetry properties required by the Liquid Drop Model prescription for the potential energy \cite{Bohr1}. This results from the fact that the harmonic oscillator potential (\ref{gamapot}) is obtained from the second order approximation of the symmetry obeying potential $a^{2}(1-\cos{3\gamma})$. Secondly, the corresponding Hamiltonian is no longer hermitical with respect to the scalar product defined with $|\sin3\gamma|d\gamma$ measure \cite{Raghe}. However, the adopted approximation was demonstrated in \cite{Caprio} to be reliable for high stiffness, condition which is achieved in the present model as will be shown in the numerical application.

In what concerns the $\beta$ degree of freedom, one will consider here an anharmonic behaviour reflected into a square well shape of the potential:
\begin{equation}
u(\beta)=
\left\{\begin{array}{l}
0,\,\,\beta\leqslant\beta_{W},\\
\infty,\,\,\beta>\beta_{W},
\end{array}\right.
\end{equation}
where $\beta_{W}$ indicates the position of the infinite wall. In this way, by varying the rigidity parameter from $\chi=0$ to $\chi\rightarrow1$, the system will change from ES-$X(5)$ model \cite{BonX5ES} to a quasi-$\gamma$-rigid solution whose $\gamma$-rigid limit is the $X(3)$ model \cite{Bon5}. Now, equation (\ref{b}) can be brought to a Bessel differential equation by the change of variable $\xi(\beta)=\beta^{\chi-\frac{3}{2}}f(\beta)$:
\begin{equation}
\left[\frac{\partial^{2}}{\partial{\beta^{2}}}+\frac{1}{\beta}\frac{\partial}{\partial{\beta}}+\left(k^{2}-\frac{\nu^{2}}{\beta^{2}}\right)\right]f(\beta)=0,
\end{equation}
where
\begin{equation}
\nu=\left[\frac{L(L+1)-(1-\chi)K^{2}}{3}+\left(\frac{3}{2}-\chi\right)^{2}+(1-\chi)3a(n_{\gamma}+1)\right]^{\frac{1}{2}}.
\end{equation}
The boundary condition $f(\beta_{W})=0$ gives the $\beta$ energy spectrum in terms of the $s$-th zero $x_{s,\nu}$ of the Bessel function $J_{\nu}(x_{s,\nu}\beta/\beta_{W})$:
\begin{equation}
\epsilon_{L,K,s,n_{\gamma}}(\beta_{W})=\left(\frac{x_{s,\nu}}{\beta_{W}}\right)^{2}.
\end{equation}
The order of the Bessel function's zero is related to the $\beta$ vibration quantum number by $n_{\beta}=s-1$. The excitation energy of the whole system in respect to the ground state is then defined as
\begin{equation}
E_{K,n_{\beta},n_{\gamma}}(L)=\frac{\hbar^{2}}{2B}\left[\epsilon_{L,K,n_{\beta}+1,n_{\gamma}}(\beta_{W})-\epsilon_{0,0,1,0}(\beta_{W})\right],
\label{et}
\end{equation}
where the set of quantum numbers $\{K,n_{\beta},n_{\gamma}\}$ uniquely identify a rotational band, {\it i.e.} $\{000\}$ for the ground band, $\{010\}$ and $\{201\}$ for the first $\beta$ and $\gamma$ excited bands and so on. Correspondingly the $\beta$ variable normalized wave function is then given as:
\begin{equation}
\xi_{L,K,n_{\beta},n_{\gamma}}(\beta)=N_{n_{\beta},\nu}\beta^{\chi-\frac{3}{2}}J_{\nu}(x_{n_{\beta}+1,\nu}\beta/\beta_{W}),
\end{equation}
where $N_{n_{\beta},\nu}$ is the normalization constant obtained from the condition
\begin{equation}
\int_{0}^{\beta_{W}}\left[\xi_{L,K,n_{\beta},n_{\gamma}}(\beta)\right]^{2}\beta^{4}d\beta=1.
\end{equation}

Finally, the full solution of the Hamiltonian (\ref{Ht}) after proper normalization and symmetrization reads \cite{Iachello1,BonX5ES}:
\begin{eqnarray}
\Psi_{LMKn_{\beta}n_{\gamma}}(\beta,\gamma,\Omega)&=&\xi_{L,K,n_{\beta},n_{\gamma}}(\beta)\eta_{n_{\gamma},|K|}(\gamma)\sqrt{\frac{2L+1}{16\pi^{2}(1+\delta_{K,0})}}\nonumber\\
&&\times\left[D_{MK}^{L}(\Omega)+(-)^{L}D_{M-K}^{L}(\Omega)\right].
\end{eqnarray}
With this function one can calculate the transition rates, employing the general expression for the quadrupole transition operator,
\begin{equation}
T_{\mu}^{(E2)}=t\beta\left[D_{\mu0}^{2}(\Omega)\cos{\gamma}+\frac{1}{\sqrt{2}}\left(D_{\mu2}^{2}(\Omega)+D_{\mu-2}^{2}(\Omega)\right)\sin{\gamma}\right],
\label{TE2}
\end{equation}
where $t$ is a scaling factor. The final result for the $E2$ transition probability is given in a factorized form \cite{Bijker,BonD}:
\begin{eqnarray}
B(E2;LKn_{\beta}n_{\gamma}\rightarrow L'K'n'_{\beta}n'_{\gamma})=\nonumber\\
\frac{5t^{2}}{16\pi}\left(C^{L\,\,2\,\,L'}_{KK'-KK'}B_{L'K'n'_{\beta}n'_{\gamma}}^{LKn_{\beta}n_{\gamma}}G_{K'n'_{\gamma}}^{Kn_{\gamma}}\right)^{2},
\end{eqnarray}
where $C$ is the Clebsch-Gordan coefficient dictating the angular momentum selection rules, while $B$ and $G$ are integrals over the shape variables $\beta$ and $\gamma$ with integration measures $\beta^{4}d\beta$ and respectively $\left|\sin{3\gamma}\right|d\gamma$. Note that in case of the $\Delta K=0$ transitions described by the first term of $E2$ operator (\ref{TE2}), the $\gamma$ integral reduces to the orthogonality condition for $\eta_{n_{\gamma},|K|}(\gamma)$ wave functions.

\begin{figure*}[th!]
\begin{center}
\includegraphics[clip,trim = 0mm 0mm 0mm 0mm,width=0.99\textwidth]{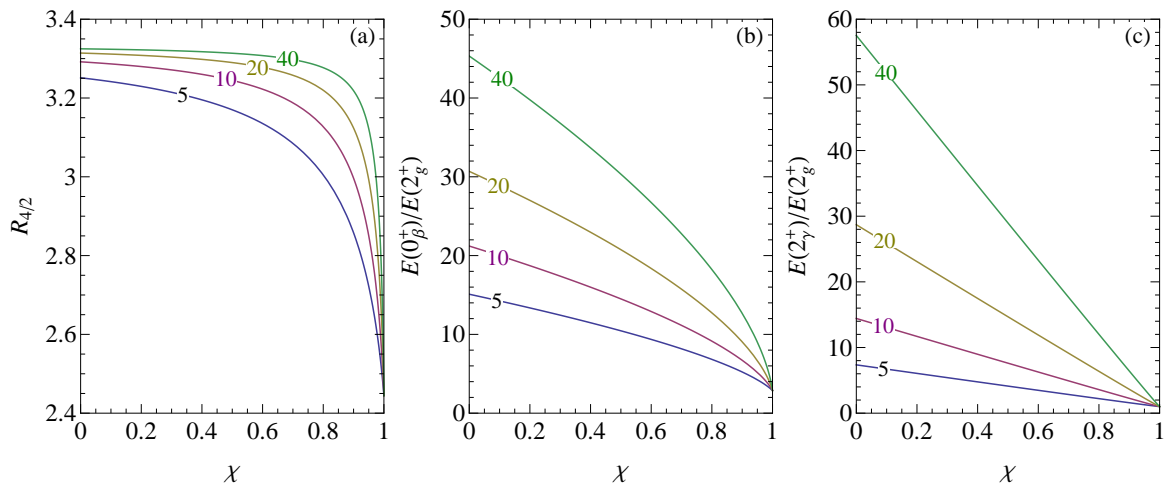}
\end{center}
\vspace{-0.2cm}
\caption{$R_{4/2}$ (a), $\beta$ (b) and $\gamma$ (c) band heads normalized to the energy of the first excited state are given as function of the rigidity parameter $\chi$ for few values of the stiffness parameter $a$, namely 5, 10, 20 and 40.}
\label{ratios}
\end{figure*}

\renewcommand{\theequation}{3.\arabic{equation}}
\section{Numerical application and discussion}
\label{sec:3}

The proposed model have two adjustable parameters not counting the scale, namely the stiffness $a$ of the $\gamma$ vibrations and the rigidity $\chi$. Indeed, considering the energy spectrum and the $E2$ transition probabilities normalized to the energy of the first excited state and respectively to its $B(E2)$ rate to the ground state, the dependence on the mass $B$ and the position of the infinite wall $\beta_{W}$ vanishes. The evolution as function of $\chi$ and $a$ of theoretically evaluated spectral observables such as $R_{4/2}=E(4_{g}^{+})/E(2_{g}^{+})$ ratio and the $\beta$ and $\gamma$ band heads normalized to the energy of the first excited state is visualised schematically in figure \ref{ratios}. From this figure one can see that the first two sets of curves corresponding to $R_{4/2}$ and $E(0_{\beta}^{+})/E(2_{g}^{+})$ collapse at $\chi\rightarrow1$ to the values of the $X(3)$ energy spectrum, namely 2.44 and respectively 2.87. Although the $X(3)$ solution by construction excludes the existence of the $\gamma$ band, our approach at $\chi\rightarrow1$ provides the value 1 for the $\gamma$ band head. At $\chi=0$, the corresponding values of ES-$X(5)$ model are recovered for all three ratios. However, up to the $\chi\rightarrow1$ limiting case the ratio $R_{4/2}$ as well as the two excited band heads have quite different behaviours. Indeed, the functions are not linear but becoming more straighter for $\beta$, and even more for the $\gamma$ band head. Another difference between the behaviour of the $R_{4/2}$ ratio and that of the excited band heads is the distinct effect of the stiffness parameter $a$. The increase in stiffness for $R_{4/2}$ is reflected in a longer and more horizontal plateau defined up to the turning point above which the line goes abruptly to the $X(3)$ value. As a matter of fact the height of the plateau seems to reach a certain saturation at higher values of the stiffness, the saturation maximal height being $R_{4/2}=3.33$ which is the $SU(3)$ rotational limit. In contradistinction, the effect of the stiffness increase on the excited band heads produce an almost proportional increase for the corresponding function.

\begin{figure}[th!]
\begin{center}
\includegraphics[clip,trim = 0mm 0mm 0mm 0mm,width=0.8\textwidth]{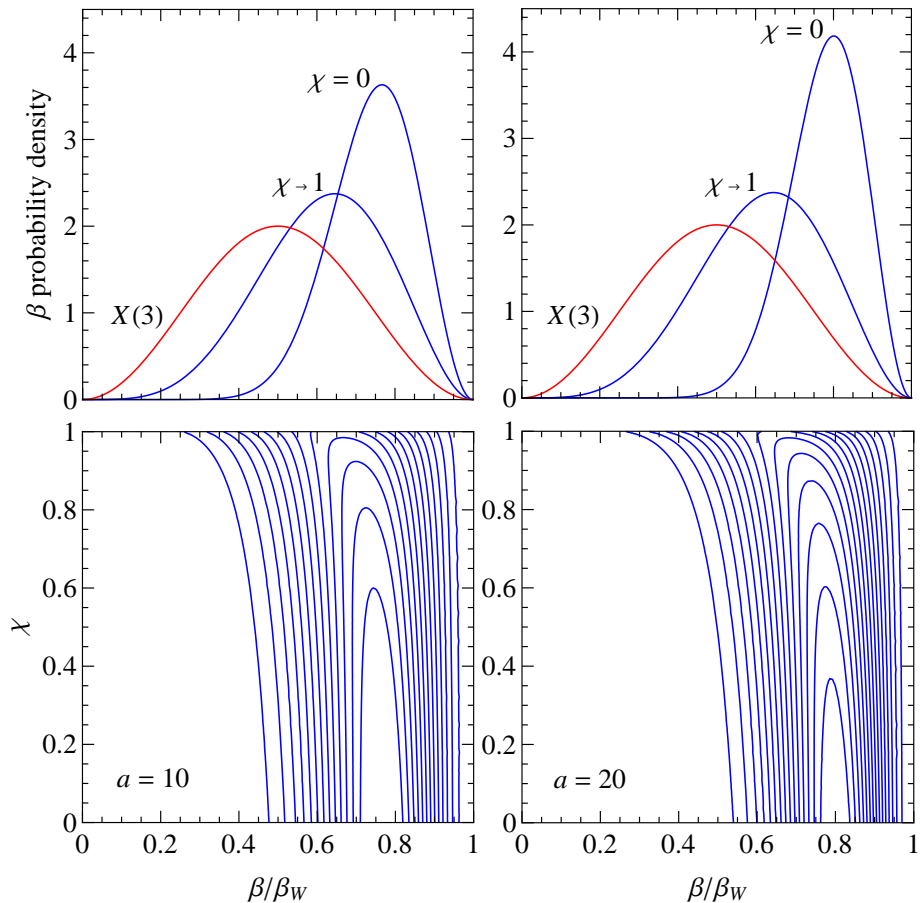}
\end{center}
\vspace{-0.2cm}
\caption{The ground state $\beta$ probability density in respect to the $d\beta$ integration measure in case of the limiting situations $\chi=0$ and $\chi\rightarrow1$ for $a=10$ and 20. The same quantity provided by the $X(3)$ model is also visualized for reference. Lines of constant probability density $\left[\xi_{0,0,0,0}(\beta)\right]^{2}\beta^{4}$ drawn as function of $\chi$ and $\beta/\beta_{W}$ show its behaviour in between the above mentioned limiting cases.}
\label{prob}
\end{figure}

In what concerns the model's wave functions, these are identical with those of ES-$X(5)$ and $X(3)$ when $\chi=0$ and $\chi\rightarrow1$, respectively. But in the $\chi\rightarrow1$ case the norm of the $\beta$ state is not the same as in the $X(3)$ model due to the different metrics of the deformation space, {\it i.e.} the integration measure in the present case is $\beta^{4}d\beta$ in comparison to $\beta^{2}d\beta$ of the $X(3)$ formalism. As a consequence of this fact, a reshaping of the $\beta$ probability distribution takes place. Indeed, as can be seen from the upper panels of figure \ref{prob}, while the probability peak in the $X(3)$ case is positioned in the center of the square well, in the present model it is shifted to higher deformation and also becomes more pronounced. From the lower panels of figure \ref{prob} one observes that this behaviour is perpetuated when going to smaller values of parameter $\chi$, reaching at some point a stagnation in the position and hight of the peak which continues up to the ES-$X(5)$ limit. Increasing the stiffness, the stagnation interval diminishes and the peak at $\chi=0$ is shifted to a higher $\beta$ deformation. This is the result of the centrifugal contribution from the $\gamma$ dependent term of the potential (\ref{pot}) which compresses the wave function to higher $\beta$ values as the stiffness of the $\gamma$ vibration increases. Such a behaviour of the $\beta$ wave function is very much similar to the $\beta$-rigid case \cite{Caprio}. In conclusion, the price to be paid when transforming $\gamma$ from a simple parameter to a variable is the confinement of the ground state $\beta$ distribution probability at higher expectation values of the $\beta$ deformation.

An extensive search for nuclei which fall in the category described by the present model was made by means of model fits on rare earth and actinide isotopes with $R_{4/2}\geq2.44$ and with all three bands experimentally accounted. The quality of the fits was judged by the quantity

\begin{equation}
\sigma=\sqrt{\frac{1}{N-1}\sum_{k=1}^{N}\left[\frac{E^{Th}_{k}}{E^{Th}(2_{g}^{+})}-\frac{E^{Exp}_{k}}{E^{Exp}(2_{g}^{+})}\right]^{2}},
\label{sigma}
\end{equation}
where $E_{k}^{Th}$ is the theoretical energy defined by (\ref{et}) with $k$ indexing angular momentum states from the ground, $\beta$ and $\gamma$ bands. The experimental realization of our model was found to occur in $^{160}$Gd, $^{162}$Dy and $^{166}$Er isotopes. As can be deduced from table \ref{spec} where the results of the fits are given in comparison to the available experimental data, the best agreement is obtained for $^{160}$Gd and $^{162}$Dy nuclei. For $^{162}$Dy one obtained the highest degree of mixing between $\gamma$-rigid and $\gamma$-stable rotation-vibration, judging by the rigidity value $\chi=0.269$ which is associated with a comparable weighting for the two competing excitations. The fits for the other two nuclei provide much higher values for $\chi$, corresponding to prevalent $\gamma$-rigid behaviour. Especially high $\gamma$-rigidity is obtained in case of $^{160}$Gd, which is damped by also high stiffness. Taking into account the conclusions drawn from figure \ref{ratios}, this high value for $\gamma$ stiffness can be ascribed to near rotational behaviour in the ground band for this particular nucleus. Although the spectral observables of $^{160}$Gd are quite far from the $X(3)$ values, it can be considered as quasi-$\gamma$-rigid in virtue of the high value of the rigidity parameter. The ground and $\gamma$ bands of all nuclei are reproduced very well, with the $\beta$ band being the principal source of the discrepancy with experiment. The overestimation of the energy differences between consecutive $\beta$ band states is a common problem for the $X(5)$ related models, which can be circumvented by considering smoothed out square well shapes for the $\beta$ potential \cite{Budaca1,Budaca2,BonUX,Rabug}. Using a polynomial potential in the $\beta$ variable not only will be more natural from the geometrical symmetry perspective but will also attenuate the compression of the wave function against the outer potential wall which in our model is partially managed by increasing the rigidity $\chi$.

In table \ref{tranz} we compare the theoretical predictions for the $E2$ transition rates computed with the values of $\chi$ and $a$ from table \ref{spec} with few commonly available experimental data, which amount to ground-ground and $\gamma$-ground transitions. In this way one can ascertain the agreement with experiment for both $\Delta K=0$ and $\Delta K=2$ transition probabilities. In the first case the agreement is very good even for high angular momentum states, while in later case the experimental values are slightly underestimated. The fact that the theoretical $B(E2)$ values vary very little between the considered nuclei is consistent with the presence of a stagnation interval in the $\beta$ transition probability density as function of $\chi$ (see figure \ref{prob}). It is instructive to compare our results also with the rigid rotor estimation of the $E2$ transition probability  \cite{Bohr3}:
\begin{equation}
B(E2;LK\,\rightarrow\,L'K')\sim\left(C^{L\,\,2\,\,L'}_{KK'-KK'}\right)^{2}.
\end{equation}
Thus, the ground-ground transitions in the present calculations are overestimated in respect to the rigid rotor results, while the $\gamma$-ground ones are almost identical. The last result is related to the quasi-rigid description of the $\gamma$ degree of freedom.

Another common trait of the considered nuclei is the nearly rigid rotor values of the experimental $R_{4/2}$ ratio which is consistent with their high experimental quadrupole moments \cite{Stone} and microscopic mean-field calculations \cite{Hilaire,Chen} which also predict very sharp probability distributions for the $\beta$ deformation of these nuclei. This is actually one of the reasons why these nuclei were found to be the best representatives of the proposed model. Indeed, the approximate $\beta$-rigid behaviour of the model due to the specific choice of the separated potential is in agreement with the experimental observations and the microscopic predictions regarding the ground state properties of these nuclei. Another aspect which pleads in favor of the present description are the high values obtained for the stiffness parameter for all three nuclei, and especially for those closer to rigidity, which also validate the small angles approximation made here \cite{Caprio,BonX5ES}. The considered nuclei present however deviations from the rotational behaviour in the ground band when going to higher spin states. Moreover, the $\beta$ and $\gamma$ bands are not even remotely close to each other as it happens in the $SU(3)$ case where these bands are degenerated. The experimental $E2$ transition probabilities for these three nuclei also deviate from the rigid rotor estimation as can be seen from table \ref{tranz}. The microscopic calculations of \cite{Hilaire} also evidenced the so called centrifugal stretching phenomenon, {\it i.e.} the increase of the average deformation with the angular momentum. As a mater of fact, our model is able to describe this property due to the presence of the centrifugal term of the potential coming from the $\gamma$ variable. This contribution shifts the inner wall of the $\beta$ square well potential, which in combination with the usual centrifugal term is responsible for the centrifugal stretching. A simplified picture of this mechanism is described in \cite{Pietrala1} with the confined $\beta$-soft model \cite{Pietrala2}. It must be mentioned here that the infinite square well shape adopted in our formalism is understood in the present context as a maximally anharmonic potential. Its usefulness is justified also by the microscopic analysis \cite{Chen} regarding the $0^{+}_{2}$ excited states in the isotopic chains corresponding to the presently discussed nuclei. The results of \cite{Chen} point to the increase in $\beta$ anharmonicity as function of the neutron number, such that its stronger for the more deformed nuclei. It is interesting that the aforementioned study suggest that the highest $\beta$ anharmonicity corresponds to the nucleus $^{160}$Gd which happens to be the best experimental realisation of the present model.

\setlength{\tabcolsep}{9.7pt}
\begin{table}[ph!]
\caption{Theoretical results for ground, $\beta$ and $\gamma$ bands energies normalized to the energy of the first excited state $2_{g}^{+}$ are compared with the available experimental data for $^{160}$Gd\cite{160Gd}, $^{162}$Dy\cite{162Dy} and $^{166}$Er\cite{166Er}. The values in parentheses denote states with uncertain band assignment and therefore were not taken into account in the fitting procedure. The adimensional parameters $\chi$ and $a$ are also given together with the corresponding deviation $\sigma$ defined by (\ref{sigma}).}
\label{spec}
\begin{center}
{\linespread{0.83}
\footnotesize
\begin{tabular}{ccccccc}
\hline\noalign{\smallskip}
&\multicolumn{2}{c}{$^{160}$Gd}&\multicolumn{2}{c}{$^{162}$Dy}&\multicolumn{2}{c}{$^{166}$Er}\\
\noalign{\smallskip}\hline\noalign{\smallskip}
$L$&Exp.&Th.&Exp.&Th.&Exp.&Th.\\
\noalign{\smallskip}\hline\noalign{\smallskip}
 $2_{g}^{+}$&   1.00 & 1.00 & 1.00 &  1.00 &  1.00 & 1.00\\
 $4_{g}^{+}$&   3.30 & 3.28 & 3.29 &  3.28 &  3.29 & 3.26\\
 $6_{g}^{+}$&   6.84 & 6.74 & 6.80 &  6.72 &  6.77 & 6.65\\
 $8_{g}^{+}$&  11.53 &11.30 &11.42 & 11.23 & 11.31 &11.06\\
 $10_{g}^{+}$& 17.28 &16.85 &17.05 & 16.72 & 16.75 &16.39\\
 $12_{g}^{+}$& 24.00 &23.33 &23.57 & 23.11 & 22.92 &22.57\\
 $14_{g}^{+}$& 31.59 &30.70 &30.89 & 30.37 & 29.65 &29.57\\
 $16_{g}^{+}$& 39.97 &38.91 &38.91 & 38.45 & 36.83 &37.35\\
 $18_{g}^{+}$&       &47.95 &47.49 & 47.33 &(44.39)&45.89\\
 $20_{g}^{+}$&       &57.79 &56.75 & 57.00 &       &55.17\\
 $22_{g}^{+}$&       &68.41 &66.35 & 67.44 &       &65.18\\
 $24_{g}^{+}$&       &79.82 &76.28 & 78.64 &       &75.91\\
 $26_{g}^{+}$&       &91.99 &      & 90.59 &       &87.36\\
\noalign{\smallskip}\hline\noalign{\smallskip}
 $0_{\beta}^{+}$& 18.33 & 19.10  &  20.66  & 18.09 & 18.12 &16.01\\
 $2_{\beta}^{+}$& 19.08 & 20.60  &  21.43  & 19.60 & 18.97 &17.55\\
 $4_{\beta}^{+}$&       & 23.99  &  23.39  & 23.00 & 20.83 &20.98\\
 $6_{\beta}^{+}$&       & 29.07  &         & 28.08 & 23.55 &26.05\\
 $8_{\beta}^{+}$&       & 35.62  &         & 34.60 & 27.24 &32.48\\
 $10_{\beta}^{+}$&      & 43.46  &         & 42.38 &(30.77)&40.09\\
 $12_{\beta}^{+}$&      & 52.46  &         & 51.28 &(32.97)&48.76\\
\noalign{\smallskip}\hline\noalign{\smallskip}
 $2_{\gamma}^{+}$& 13.13& 13.21  &11.01& 11.07 &  9.75 & 9.59\\
 $3_{\gamma}^{+}$& 14.05& 14.09  &11.94& 11.95 & 10.67 &10.46\\
 $4_{\gamma}^{+}$& 15.25& 15.24  &13.15& 13.12 & 11.87 &11.61\\
 $5_{\gamma}^{+}$& 16.76& 16.69  &14.66& 14.56 & 13.34 &13.03\\
 $6_{\gamma}^{+}$& 18.51& 18.41  &16.42& 16.28 & 15.09 &14.71\\
 $7_{\gamma}^{+}$& 20.58& 20.38  &18.48& 18.25 & 17.08 &16.64\\
 $8_{\gamma}^{+}$& 22.81& 22.62  &20.71& 20.48 & 19.31 &18.82\\
 $9_{\gamma}^{+}$&      & 25.11  &23.28& 22.96 & 21.74 &21.23\\
 $10_{\gamma}^{+}$&28.14& 27.84  &25.88& 25.68 & 24.37 &23.87\\
 $11_{\gamma}^{+}$&     & 30.82  &28.98& 28.63 & 27.18 &26.73\\
 $12_{\gamma}^{+}$&34.31& 34.02  &32.52& 31.80 & 30.14 &29.80\\
 $13_{\gamma}^{+}$&     & 37.46  &35.45& 35.20 &(32.94)&33.09\\
 $14_{\gamma}^{+}$&     & 41.11  &39.00& 38.82 &(35.74)&36.59\\
 $15_{\gamma}^{+}$&     & 44.99  &42.57& 42.65 &       &40.28\\
 $16_{\gamma}^{+}$&     & 49.08  &46.30& 46.69 &       &44.18\\
 $17_{\gamma}^{+}$&     & 53.38  &50.08& 50.94 &       &48.27\\
 $18_{\gamma}^{+}$&     & 57.90  &53.84& 55.39 &       &52.55\\
 $19_{\gamma}^{+}$&     & 62.62  &     & 60.05 &       &57.03\\
 $20_{\gamma}^{+}$&     & 67.54  &     & 64.90 &       &61.69\\
 \noalign{\smallskip}\hline\noalign{\smallskip}
$\chi$&\multicolumn{2}{c}{0.948} &\multicolumn{2}{c}{0.269} &\multicolumn{2}{c}{0.848}\\
$a$  &\multicolumn{2}{c}{168.899}&\multicolumn{2}{c}{10.309}&\multicolumn{2}{c}{41.191}\\
$\sigma$  &\multicolumn{2}{c}{0.567}&\multicolumn{2}{c}{0.845}&\multicolumn{2}{c}{1.359}\\
\noalign{\smallskip}\hline
\end{tabular}}
\end{center}
\end{table}

\setlength{\tabcolsep}{4.3pt}
\begin{table}[th!]
\caption{Several commonly available experimental $E2$ transition probabilities for $^{160}$Gd\cite{160Gd}, $^{162}$Dy\cite{162Dy} and $^{166}$Er\cite{166Er} are confronted with the model's predictions. $\Delta K=0$ transition rates are normalized to the $2_{g}^{+}\rightarrow0^{+}_{g}$ transition, while $\Delta K=2$ transitions to the $2_{\gamma}^{+}\rightarrow0^{+}_{g}$ transition, as in \cite{BonD,Bijker}. The rigid rotor estimations are also presented for reference.}
\label{tranz}
\begin{center}
\begin{tabular}{cllllllll}
\hline\noalign{\smallskip}
Nucleus&$\frac{4^{+}_{g}\rightarrow2^{+}_{g}}{2^{+}_{g}\rightarrow0^{+}_{g}}$& $\frac{6^{+}_{g}\rightarrow4^{+}_{g}}{2^{+}_{g}\rightarrow0^{+}_{g}}$& $\frac{8^{+}_{g}\rightarrow6^{+}_{g}}{2^{+}_{g}\rightarrow0^{+}_{g}}$& $\frac{10^{+}_{g}\rightarrow8^{+}_{g}}{2^{+}_{g}\rightarrow0^{+}_{g}}$& $\frac{12^{+}_{g}\rightarrow10^{+}_{g}}{2^{+}_{g}\rightarrow0^{+}_{g}}$&
$\frac{14^{+}_{g}\rightarrow12^{+}_{g}}{2^{+}_{g}\rightarrow0^{+}_{g}}$& $\frac{2^{+}_{\gamma}\rightarrow2^{+}_{g}}{2^{+}_{\gamma}\rightarrow0^{+}_{g}}$& $\frac{2^{+}_{\gamma}\rightarrow4^{+}_{g}}{2^{+}_{\gamma}\rightarrow0^{+}_{g}}$\\
\noalign{\smallskip}\hline\noalign{\smallskip}
$^{160}$Gd&&&&&&&1.87(12)&0.189(29)\\
&1.45&1.62&1.74&1.83&1.90&1.97&1.441&0.073\\
$^{162}$Dy&1.42(6)&1.48(9)&1.70(9)&1.72(11)&1.62(20)&1.62(20)&1.78(16)&0.137(12)\\
&1.45&1.64&1.77&1.87&1.96&2.04&1.444&0.074\\
$^{166}$Er&1.44(6)&1.71(10)&1.72(8)&1.80(9)&1.71(10)&1.84(23)&1.86(14)&0.151(10)\\
&1.45&1.64&1.77&1.87&1.95&2.03&1.445&0.074\\
Rigid&1.43&1.57&1.65&1.69&1.72&1.74&1.429&0.071\\
rotor&&&&&&&&\\
\noalign{\smallskip}\hline
\end{tabular}
\end{center}
\end{table}

The nuclei $^{162}$Dy and $^{166}$Er were previously reported as candidates for ES-$X(5)$ model \cite{BonX5ES} along with $^{162}$Gd and $^{176}$Yb. Except $^{162}$Dy, the nomination was based only on the $R_{4/2}$ ratio and the excited band heads. The inclusion of the $\gamma$-rigidity factor significantly improved the agreement with experiment for $^{162}$Dy regarding the whole energy spectrum. Another important point regarding the comparison with ES-$X(5)$ approach which is worth mentioning is the fact that present model's fits for $^{162}$Gd offered approximately the same results, namely $\chi\approx0$ and $a=8.5$ with $\sigma=0.415$ against $a($ES-$X(5))=8.3$ \cite{BonX5ES}. Numerical applications on the nuclei considered here were also performed using the exactly separable version of Bohr Hamiltonian with a Davidson potential \cite{BonD}, whose accord with experiment is poorer for $^{162}$Dy and $^{160}$Gd and better for $^{166}$Er. However, it must be mentioned that the experimental $\beta$ band considered in \cite{BonD} for the $^{162}$Dy nucleus, is presently identified as S-band \cite{162Dy}. The experimental $\beta$ band for $^{160}$Gd in our comparison is also different as per suggestion of \cite{160Gd}. A better description of these three nuclei was obtained with a Morse potential \cite{Morse}, as well as with Davidson \cite{Bon2} and Kratzer \cite{Bon3} potentials combined with a deformation dependent mass term. However all these models posses an additional free parameter.

\section{Conclusions}

A simple exactly separable model was constructed by taking the kinetic energy of the Bohr Hamiltonian as a combination of prolate $\gamma$-rigid and $\gamma$-stable rotation-vibration kinetic operators. The relative weight of these two components is managed through a so called rigidity parameter $\chi$. It also defines the potential, such that when $\chi=0$ the potential takes the separated form $u(\beta)+u(\gamma)/\beta^{2}$, while for $\chi\rightarrow1$ the $\gamma$ potential disappears. The $\gamma$ and angular part of the eigenvalue problem associated to such a model is treated as in \cite{BonX5ES}, {\it i.e.} by applying a small angle approximation for the $\gamma$ trigonometric functions and the corresponding potential. The later acquires a harmonic oscillator form centered on $\gamma=0$ point and characterized by a stiffness parameter $a$. As the scope of the present approach was to insert the rigidity degree of freedom into the description of critical point collective motion, an infinite square well shape was chosen for the $\beta$ potential. This choice is consistent with the requirement that in a critical point of a shape phase transition the potential must be flat or having multiple degenerated minima \cite{Casten}. With this choice, the $\beta$ part of the problem is brought to a Bessel type equation. As a result, the model's energy spectrum and the $E2$ transition probabilities depend on two parameters, excepting the scale. Numerical applications showed that the energy spectrum at $\chi\rightarrow1$ recovers the $X(3)$ values regardless of the stiffness value. However, in the rest of the interval, the stiffness plays an important role in defining the excited band heads which are rising proportionally. On the other hand, its effect on the $R_{4/2}$ ratio has a saturation behaviour, tending to the rotational limit 3.33 at high values. Taking into account the above considerations, the spectral characteristics of the present model can be succinctly described as $R_{4/2}>2.44$, $E(0_{\beta}^{+})/E(2_{g}^{+})>2.87$ and $E(2_{\gamma}^{+})/E(2_{g}^{+})>1$. It is needless to say that in the lower limit of the rigidity at $\chi=0$ the energy spectrum recovers the corresponding ES-$X(5)$ structure.

Although there are two free parameters, the present solution is as poor in candidate nuclei as $X(5)$ critical point symmetry and its exactly solvable analog ES-$X(5)$. This is actually a common trait for parameter independent critical point solutions. Indeed, because the energy spectrum of nuclei change discretely with $N$ and $Z$, the critical point of phase transition can very well be missed in between two nuclei. Moreover, the present model is based on the irrotational flow interpretation of the moments of inertia while it is well known \cite{Bohr3} that the actual moments of inertia lie in between these hydrodynamical realizations and the rigid ones. As a result, the search for model candidates revealed only three nuclei with good agreement with experiment and a high degree of mixing between $\gamma$ rigidity and stability. These are the $N=96$ isotones $^{160}$Gd and $^{162}$Dy together with $^{166}$Er nucleus which have 98 neutrons. Considering the structure of the proposed model it is then not surprising to find its experimental realisations in the heavier and more deformed branches of isotopic chains undergoing shape phase transitions. The transitional region to which they belong is also known to exhibit level sequences of the collective spectrum with alternating parities which are understood as a consequence of the quadrupole-octupole coherent coupling \cite{Minkov}. Such that for a more realistic description of these nuclei the inclusion of at least octupole degrees of freedom is prerequisite. Although the hexadecapole deformation induce irremediable complications it must not be completely neglected given the open possibility for the realization of octahedral discreet symmetry in even-even rare earth nuclei along with the tetrahedral one \cite{Dudek}.

In what concerns numerical results, the fits for $^{162}$Dy designate it more closer to the $\gamma$-stable behavior, while for the other two the rigid excitations are stronger. Moreover, $^{160}$Gd seems to be falling into the quasi-$\gamma$-rigid category. Also our model fit for the $^{162}$Gd nucleus, confirms it as a ES-$X(5)$ representative \cite{BonX5ES} which in the present formalism amounts to $\chi=0$. Regarding the electromagnetic properties of these nuclei, theoretical estimations show a good agreement with the available experimental $B(E2)$ values for $\Delta K=0$ as well as $\Delta K=2$ transitions. In both cases, one observes a very weak dependence on the two parameters.

Because of the above-mentioned aspects one may say that the hybrid formalism proposed in this paper unveils alternative features of the collective motion in the vicinity of the critical point of a spherical to deformed shape phase transition. The combining scheme described in the previous sections can be extended also to quadrupole excitations closer to the critical point by considering a different $\beta$ potential, such as for example harmonic oscillator or Davidson \cite{Davidson} potentials. One expects that the predictive power of the model would increase based on a greater number of treatable nuclei with such potentials \cite{BonX5ES,BonD}.

\ack
The authors acknowledges the financial support received from the Romanian Ministry of Education and Research, through the Project PN-09-37-01-02/2013.

\end{document}